\documentclass[aps,pre,onecolumn,showpacs,showkeys,a4paper]{revtex4}
\input epsf  
\usepackage{graphicx} 
 \usepackage{amsmath}
  \usepackage{amssymb}
  \usepackage{amscd}
  \usepackage{bbm}
\usepackage[all]{xy}

\begin{document} 

 \title{Exact results for  an asymmetric annihilation process
 with open boundaries}
\author{Arvind Ayyer}
  \email{arvind.ayyer@cea.fr}
  \affiliation{Institut  de Physique Th\'eorique, C. E. A.  Saclay,
 91191 Gif-sur-Yvette Cedex, France}
\author{Kirone Mallick}
  \email{kirone.mallick@.cea.fr}
  \affiliation{Institut  de Physique Th\'eorique, C. E. A.  Saclay,
 91191 Gif-sur-Yvette Cedex, France}
   \date{\today}



\newenvironment{proofof}[1]{\medskip\noindent
   \textbf{Proof of #1:} }{\hfill $\blacksquare$\par\medskip}

\newcommand{\be}{\begin{equation}}
\newcommand{\ee}{\end{equation}}
\newcommand{\ds}{\displaystyle}
\newcommand{\id}[1]{\mathbbm{1}_{#1}}
\newcommand{\iden}{\mathbbm{1}}
\newcommand{\dens}[1]{\langle \eta_{#1} \rangle}
\newcommand{\sigid}{(\sigma \otimes \iden)}
\newcommand{\exi}[1]{\langle \xi_{#1} \rangle}
\newcommand{\aver}[1]{\langle #1 \rangle}

\newtheorem{thm}{Theorem}
\newtheorem{cor}[thm]{Corollary}
\newtheorem{lem}[thm]{Lemma}
\newtheorem{defn}{Definition}
\newtheorem{rem}{Remark}
\newtheorem{conj}{Conjecture}

\makeatletter
\def\Ddots{\mathinner{\mkern1mu\raise\p@
\vbox{\kern7\p@\hbox{.}}\mkern2mu
\raise4\p@\hbox{.}\mkern2mu\raise7\p@\hbox{.}\mkern1mu}}
\makeatother

\newcommand{\marginnote}[1]
           {\mbox{}\marginpar{\raggedright\hspace{0pt}{$\blacktriangleright$
#1}}}

\begin{abstract}
  We consider a  nonequilibrium reaction-diffusion model on a
  finite one dimensional lattice with bulk and boundary dynamics
  inspired by Glauber dynamics of the Ising model. We show that the
  model has a rich algebraic structure that we use to calculate its
  properties.  In particular, we show that the Markov dynamics for a
  system of a given size can be embedded in the dynamics of systems of
  higher sizes.  This remark leads us to devise a technique  we
  call the {\em transfer matrix Ansatz} that allows us to determine
  the steady state distribution and correlation functions.
  Furthermore, we show that the disorder variables satisfy very simple
  properties and we give a conjecture for the characteristic
  polynomial of Markov matrices. Lastly, we compare the transfer
  matrix Ansatz used here with the matrix product representation of
  the steady state of one-dimensional stochastic models.

\end{abstract}

\maketitle

  \section{Introduction} 
  \label{sec:intro}

  The study of systems far from equilibrium has been greatly helped by
  the discovery of exactly solvable models, because explicit
  computations for these models provide us with excellent testing
  grounds for general hypotheses about nonequilibrium statistical
  mechanics \cite{spohn,zia}.
 
 One important difference between equilibrium and nonequilibrium
 behaviour is encoded in the detailed balance condition. This
 condition states that at equilibrium the total transition rate
 between two arbitrary micro-configurations vanishes identically
 \cite{vankampen}.  Conversely, nonequilibrium steady states usually
 break detailed balance, which results in the existence of current
 loops in the configuration space of the system, leading to a non-zero
 macroscopic (physical) current that transports matter, momentum or
 energy from one region of the system to another.  Because of this
 current, the boundaries of the system can affect its bulk and the
 modification of the boundary conditions through a control parameter
 can induce dynamical phase transitions even in one dimensional
 systems \cite{krug}.  Such  sensitiveness to the boundary conditions
 is well demonstrated by exact results obtained for the asymmetric
 exclusion process (ASEP) in one dimension, which is one of the
 simplest examples of a driven lattice gas and one of the most
 exhaustively investigated interacting particle systems
 \cite{derridareview,schutzrev,liggett}.  For the ASEP on a periodic
 ring the steady state is uniform and all configurations have the same
 stationary weight; in contrast, for the ASEP on a finite lattice with
 open boundary conditions (that allow injection and removal of
 particles at the end sites) the measure is non-uniform and in the
 limit of large sizes, the system can exist in three different phases:
 maximal current, low density and high density (the last two phases
 being separated by a line of shocks).  The exact expression of the
 stationary measure valid for any system size was first derived in
 \cite{DEHP}, introducing a method now called the matrix product
 representation that has become an important technique for one
 dimensional interacting particle processes.  In particular, this
 matrix representation has an interpretation in terms of discrete
 lattice paths that leads to exact combinatorial results for finite
 size systems (see \cite{MartinRev} for an exhaustive and recent
 review).  Thanks to this matrix product method and to more standard
 integrability techniques such as the Bethe Ansatz \cite{ogkmrev}, an
 plethora of results have been derived for the ASEP
 \cite{derjstat}.
 
 Here, we study a reaction-diffusion model on a finite lattice of $L$
 sites with open boundaries in which hard-core particles perform
 asymmetric jumps and can undergo pairwise annihilation.  This model
 is inspired by the Glauber dynamics \cite{Glauber} for the Ising
 model, in which the elementary excitations (or particles) are not the
 individual spins but rather the domain walls between sets of opposite
 spins.  More precisely, we shall consider the totally asymmetric
 Glauber dynamics where each spin changes its orientation with a
 certain probability based strictly on the spin to its left. Hence,
 the domain walls, which are represented by particles, move only to
 the right and, if two of them collide, they annihilate each other.
 Reaction-diffusion models have been thoroughly studied in the
 nonequilibrium statistical physics literature (see
 e.g. \cite{ElskensFrisch,AvrahamDoering} for similar studies on the real line
 and \cite{Racz,Sid,PaulK1, PaulK2, Hinrichsenetal} on the lattice)
 and their relation to non-Hermitian spin chains has been established
 \cite{Lushnikov, Barma}.  In particular, it has been shown that for
 certain values of the reaction-rates, they reduce to free fermion
 models that can be solved by using Jordan-Wigner fermionization
 techniques
 \cite{Stinchcombe1,Stinchcombe2,Schutz,deOliveira,Sasaki,Mobilia,Paessens}.
 This method was applied to periodic boundary conditions where
 translation invariance allows the use of Fourier transform to
 diagonalize the free fermion Hamiltonian.  More recently, similar
 techniques were used by Farago \cite{Farago}, and Farago and Pitard
 \cite{Pitard} to calculate large deviation functions of the time
 integrated injected power when a single spin is allowed to perform
 Poissonian flips.
 
 We shall study the effect of open boundaries conditions on the long
 time dynamics of the model.  At the boundary sites of the lattice, we
 allow creation and destruction of particles in such a manner that the
 boundary dynamics is compatible with the bulk dynamics.  Here, the
 Jordan-Wigner transformation introduces non-local terms in the free
 fermion Hamiltonian and the lack of translation invariance precludes
 the use of the Fourier transform.
   
 We shall define a one-parameter family of models with a nontrivial
 steady state that generalizes the totally asymmetric exclusion
 process (TASEP) on a finite lattice, first solved in \cite{DEHP}. For
 a special value of the parameter (which corresponds to the free
 fermion point), we calculate the exact nonequilibrium steady state
 measure by using a recursion between systems of different sizes.
 Here, the key to the solution is not a matrix product representation
 but a  different approach: using a linear transform between
 configuration spaces of different dimensions, we shall prove that the
 Markov matrix of the system of size $L$ can be embedded in the Markov
 matrix of the system of size $L+1$. The existence of such a
 rectangular (thus, non-invertible) similarity matrix, that will be
 called a `transfer matrix Ansatz', will allow us to derive exact
 combinatorial expressions for the local density and for correlation
 functions of the model, and to extract their asymptotic behaviour in
 the infinite system size limit.

 The organization of the paper is as follows: In
 Section~\ref{sec:mod}, we introduce the model. In
 Section~\ref{sec:tma}, we describe the transfer matrix Ansatz in
 general terms and apply it to our system. This leads us to a closed
 form expression for the `partition function' that normalizes the
 stationary probabilities and we derive a formula for the joint
 occupation of the first $k$ sites for all $k$. In
 Section~\ref{sec:jw}, we define the disorder variables and calculate
 the associated one-point and two-point functions. We use these variables to
 calculate the density and the rate of evaporation; we also prove some
 general properties  of higher correlation functions. In
 Section~\ref{sec:mm}, we conjecture some properties of the spectrum
 of the Markov matrices.  Section~\ref{sec:con} is devoted to
 concluding remarks and some  open problems.  In Appendix~\ref{sec:rec}, we
 write explicit recursion relations for the steady state probabilities
 that allow us to compare the transfer matrix Ansatz with the matrix
 product representation.

\section{The Model}
 \label{sec:mod}

We consider a  nonequilibrium system on a finite lattice with 
 $L$ sites labelled from 1 to $L$. We denote the
boundary of the domain wall, that is the boundary between oppositely
charged spins, by a particle using the standard notation
 $1.$ Empty sites are denoted by 0. The evolution rule in the
  bulk, biased
 Glauber dynamics,  is thus  given by 
\be \label{bulk}
\begin{split}
&10 \to 01 \; \text{with rate } 1 \, , \\
&11 \to 00 \; \text{with rate } \lambda \, .
\end{split}
\ee We remark that the first rule represents the movement of the
 domain wall to the right (eg. $++ \big\vert -- \to +++ \big\vert -$)
 and the second, the annihilation of two domain walls (eg. $++
 \big\vert - \big\vert + \to ++++$).

 The evolution of the first  site is  given by
\be \label{lbound}
\begin{split}
&0 \to 1 \; \text{with rate } \alpha,\\
&1 \to 0 \; \text{with rate } \alpha \lambda. 
\end{split}
\ee

 Particles can exit from the last site according to 
\be \label{rbound}
1 \to 0 \; \text{with rate } \beta.
\ee
 The rules \eqref{lbound} and  \eqref{rbound} were constructed
 by considering   the finite lattice  to be  a   `window' 
 of  the infinite one-dimensional lattice. Suppose 
 that  there is a virtual  site labelled 0 to the left   of the first site
  and a   virtual  site labelled $L+1$  to the right  of the last site. 
  The  left  boundary conditions are deduced from the bulk rules \eqref{bulk} 
 by looking  at the  second component of the bond $0-1$  and the right  boundary 
 conditions 
 are obtained  by looking at the first component  of the bond $L-(L+1)$; 
  $\alpha$ and $\beta$  are free control parameters.

The rules \eqref{bulk}, \eqref{lbound} and  \eqref{rbound} fully
 define the nonequilibrium model by allowing one to construct
 explicitly its  Markov matrix of dimension  $2^L$.
Notice that we  use terminology directly from the original TASEP
problem for the bulk and boundary rates and that we have ensured the
consistency of the bulk and the  boundary rates by fixing the
rates appropriately. Indeed, the value $\lambda =0$ corresponds
 exactly to the  TASEP solved in \cite{DEHP}. 

  The model can be discussed    using an approximate 
mean field  argument. The evolution  equation for the
density at site $i>1$ is given by
\be
 \frac d{dt} \aver{\eta_i} =  \aver{\eta_{i-1}(1-(\lambda +1)\eta_i) }
-\aver{\eta_i(1-(\lambda -1)\eta_{i+1})  } \, , 
\ee
 where $\eta_i$  represents  the
  occupation variable at site $i$.
 For $\lambda =1$,  these equations become simpler because   the state
 of a site  depends only  on  the preceding sites. Then, 
  using the mean-field assumption,  we obtain  the  following recursion
  valid in  the stationary state:
\be
\aver{\eta_i} = \frac{\aver{\eta_{i-1}}}{1+2\aver{\eta_{i-1}}}.
\ee
The stationary  density at the first site can be obtained  exactly by writing the
equation
\be
0 = \frac d{dt} \aver{\eta_1} =  \alpha \aver{(1-2\eta_i) }
-\aver{\eta_1},
\ee
and therefore one obtains the general formula,
\be
\aver{\eta_k} = \frac{\alpha}{1+2k\alpha}.
\ee
This implies that the density falls off like $k^{-1}$ for large $k$.
This mean-field result  is wrong: in one-dimension, 
 the actual exponent is  $-1/2$. Note  that
the exit rate $\beta$ did not enter the calculation.

 The fact that the value $\lambda=1$  plays a special  role can readily
 be understood from the dynamical rules: indeed 
  the exit rate from any site $i$ is equal to 1
  whether site $i+1$ is occupied or not. Besides,  $\lambda=1$
 corresponds to the free fermion point of the associated
 spin chain (see Section~\ref{sec:con} for a more  detailed explanation 
  of this fact).

\begin{rem} \label{rem:corrfn}
  More generally, for  $\lambda=1$,  this model has 
 the following important property. The correlation function $
\langle \eta_{i_1} \dots \eta_{i_n} \rangle$   does not depend on the
state of $\eta_{i_n+1}$ simply because the exit rate from any
configuration that contributes to this expectation value is equal to 1 
whether site $i_n+1$ is occupied or not (by \eqref{bulk}).
In particular, for a system of size $L$, all correlation functions
which depend on sites other than the last one are going to be strictly
independent of $\beta$ and moreover, any correlation function of the
form $\langle \eta_{i_1} \dots \eta_{i_n} \rangle$ is independent of
$L$ as long as $L \geq i_n+1$.
\end{rem}

 For the rest of the paper, we will take
$\lambda=1$ as that is the only case for  which  we can derive explicit
 combinatorial formulae.

\section{Semi-similarity between  Markov processes} \label{sec:tma}

\subsection{The Transfer Matrix Ansatz}
Let $M_L$ denote the Markov matrix for a system with $L$
sites. Typically the size of the matrix will be exponential in $L$. 
 We first give a general
definition of the ``transfer matrix Ansatz'' and then apply it to the
specific case of the problem defined above.

Let us consider a family of Markov processes defined by Markov
matrices $\{M_L \}$ of increasing 
sizes (in  most physical applications, $L$ is  the size of the system).
We shall see that for certain systems there exists a natural
 embedding of  $M_L$  into  $M_{L+1}$.

\begin{defn}
We say that a  family of  Markov processes  satisfies the {\em Transfer Matrix
  Ansatz} if there exist matrices $T_{L,L+1}$ for all sizes $L$
such that 
\be \label{tm}
M_{L+1} T_{L,L+1} = T_{L,L+1} M_L \, .
\ee
 We also  impose that  this equality is nontrivial in the sense that
\be
M_{L+1} T_{L,L+1} \neq 0. \label{cond:nontriv}
\ee
\end{defn}
   The  rectangular transfer matrices $T_{L,L+1}$ can be interpreted as {\em
  semi-similarity transformations} connecting Markov matrices of different
sizes. Another way to view the transfer matrices is that
that the following diagram
\be
   \xymatrix{ \Omega_L \ar[r]^{M_L} \ar[d]_{T_{L,L+1}} & 
     \Omega_L \ar[d]^{T_{L,L+1}} \\
     \Omega_{L+1} \ar[r]^{M_{L+1}} & \Omega_{L+1} }
\ee
commutes, where $\Omega_L$ is the space of $2^L$ configurations of size $L$.

 We first explain the importance of the  last  condition. 
For the nonequilibrium systems that we are interested
in, there is usually one unique steady state, which means the
multiplicity of the zero eigenvalue is one. If  $|v_L\rangle$
 is a non-zero  vector in the kernel  of $M_L$ and 
 $\langle 1_L| = (1,1,\ldots,1)$,  
   the matrix $V_{L,L+1} =
|v_{L+1}\rangle\langle 1_L| $ satisfies  \eqref{tm} 
since the Markov matrices satisfy the condition
$\langle 1_L| M_L = 0$. However, we have 
$ M_{L+1} V_{L,L+1} = V_{L,L+1} M_L=0,$ 
 and  the condition \eqref{cond:nontriv} is  violated:
therefore this  trivial solution is excluded.

 The above definition   leads immediately to some important 
observations.  First we have
\be
 \label{constructvL}
0 = T_{L,L+1} M_L |v_L \rangle = M_{L+1} T_{L,L+1} |v_L \rangle,
\ee
which, assuming $T_{L,L+1} |v_L \rangle \neq 0$, and taking
 into account the uniqueness of
the steady state, allows us to define $|v_{L+1} \rangle$ so that 
\be \label{nontriv}
T_{L,L+1} |v_L \rangle = |v_{L+1} \rangle.
\ee
The other important consequence  is related to the
eigenvalues of $M_L$. Let $|w \rangle$ be any eigenvector of $M_L$ with
eigenvalue $\mu$. Then
\be
\mu T_{L,L+1} |w \rangle = T_{L,L+1} M_L |w \rangle = M_{L+1}
T_{L,L+1} |w \rangle,
\ee
which, again  assuming $T_{L,L+1} |w \rangle \neq 0$, means that $T_{L,L+1}
|w \rangle$ is an eigenvector of $M_{L+1}$ with the same 
eigenvalue $\mu$. In other words, an  eigenvector   of $M_L$ 
 that is not in the kernel $ T_{L,L+1}$   is  also an 
eigenvector of $M_{L+1}$ with the same eigenvalue 
 (if the matrix $M_L$ is not
diagonalizable, then we cannot immediately make a statement about the
multiplicities). Conversely, if $M_L$  and
$M_{L+1}$ have a common eigenvalue $\mu$, then the rank-one
 rectangular matrix 
$ |\mu_{L+1}\rangle\langle \mu_L| $ is a transfer matrix; 
  but  the  image of the steady state 
   $|v_L \rangle$ of  $M_L$ by this matrix 
 vanishes. Hence,   $ |\mu_{L+1}\rangle\langle \mu_L| $   cannot be used to construct
 the steady state of $M_{L+1}$ knowing  $|v_L \rangle$. 
 Therefore, in order to study the stationary
 state,  we must look  for  transfer matrices  that
 satisfy the additional condition:  $T_{L,L+1}  |v_L \rangle \neq 0$. 

\subsection{The case of the asymmetric annihilation process}

For the system we consider here, the Markov matrices are of size
$2^L$. It is most convenient for us to take the naturally ordered
basis of binary sequences of size $L$.
For example,
when $L=2$, the ordered list is $(00,01,10,11)$.
The first important  observation  is that there is a recursion of order one
among the Markov matrices.
\begin{thm} \label{thm:mm}
Let $\sigma$ denote the matrix
\be
 \sigma = \begin{pmatrix}
  0 & 1 \\
  1 & 0
  \end{pmatrix},
\ee
and $\mathbbm{1}_L$ denote the identity matrix of size $2^L$. Then 
\be \label{mmdecomp}
M_L = \left( \begin{array}{cc}
M_{L-1} - \alpha (\sigma \otimes \id{L-2}) & \alpha \id{L-1} + (\sigma
\otimes \id{L-2}) \\ 
\alpha \id{L-1} & M_{L-1} - \id{L-1} - \alpha (\sigma \otimes \id{L-2} )
\end{array} \right),
\ee
where $M_L$ is written as a $2 \times 2$ block matrix with each block
made up of matrices of size $2^{L-1}$. The initial matrix for
 $L=1$ is given
by
\be \label{mmic}
M_1 = \begin{pmatrix}
  -\alpha & \alpha+\beta \\
  \alpha & -\alpha-\beta
  \end{pmatrix}.
\ee
\end{thm}

 Equation \eqref{mmdecomp} is proved as follows. 
 Let $v,v'$ denote binary vectors of length $L$ and $w,w'$ denote
vectors of length $L-1$. The $2^L$ binary  configurations
 of the system of size $L$ are listed increasingly from 
 $00\ldots 00$  to $11\ldots 11.$
 If $v_j$  and  $v_i$  are two binary configurations 
  the $(i,j)$th entry of the Markov matrix
 $M_{L}$  is the rate of the
process $v_j \to v_i$ for  $i \neq j$ and the diagonal entries are given
by $-\sum_{j \neq i} (M_L)_{j,i}$.
Decompose the Markov matrix of size $2^L$ in four blocks of size
$2^{L-1}$ according to the first bit,
\be 
M_L = \left( \begin{array}{cc}
M_{00} & M_{01}\\
M_{10} & M_{11}
\end{array} \right).
\ee

 First let us consider $M_{10}$. This encodes all transitions $v=0w
\to v'=1w'$. The only allowed transition is $0w \to 1w$ with rate
$\alpha$. Hence $M_{10} = \alpha \id{L-1}$. Similarly, $M_{00}$
encodes transition of the form $v=0w \to v'=0w'$. If $w \neq w'$ and
the first bits  of both $w$ and $w'$ are the same, the transition is
either in the bulk or on the right boundary, which is encoded
completely in the matrix for $w  \to w'$, namely $M_{L-1}$. There is, however, 
an additional transition which is present in the $L-1$ system but  is
not present in the $L$ system. Suppose
 that  the first bit $w_1$  of $w$ 
 and the first bit $w'_1$  of $w'$  satisfy $w_1 = 1 - w'_1$  and 
 that  all the
remaining bits are the same, then $w \to w'$ with rate $\alpha$
because of \eqref{lbound}. We therefore 
have to subtract these transitions from $M_{00}$. Thus  $M_{00} =
M_{L-1}-\alpha (\sigma \otimes \id{L-2})$.
 Note  that the sum 
of the elements of any  of the first $2^{L-1}$ columns of $M_L$ is indeed 
equal to zero, because in each column 
 we have once  added  and once subtracted $\alpha$  and
therefore the diagonal terms are unchanged.

For $M_{01}$, we consider transitions of the form $v=1w \to
v'=0w'$. If $w=w'$,  this transition occurs with rate $\alpha$. One other
transition depends on the first bits of $w,w'$ and assumes  that all the  other
bits are the same. If $w_1=1-w'_1$, then $w \to w'$ with rate 1 in the system 
of size $L$ because of the
transitions \eqref{bulk}. Thus $M_{01}=\alpha \id{L-1} + (\sigma
\otimes \id{L-2})$. Lastly $M_{11}$ encodes transitions $v=1w \to
v'=1w'$. The argument is now similar to that of  the $M_{00}$ case. If the
first bits  of $w$ and $w'$ are the same, then all the other transitions
are encoded by $M_{L-1}$. Additionally, if $w_1 = 1- w'_1$ and the
other bits are the same, then there is a transition given by
\eqref{lbound} for the system of 
size $L-1$  which is not present in the system of size $L$ and we have
to subtract this contribution. Besides, 
 in  $M_{01}$ we have  added $1+\alpha$ to each
column and we have subtracted $\alpha$ in $M_{11}$,
 therefore, to ensure probability conservation,   we must
subtract 1  from  the diagonal in $M_{11}$, which
results in $M_{11}= M_{L-1} - \id{L-1} - \alpha (\sigma \otimes \id{L-2} )$.

 This system satisfies the transfer matrix Ansatz: one special
 solution can be constructed recursively as we now explain. 
 In fact  from   the study of small systems (up to size $L=6$), 
 we found   $2^L$ independent  solutions  of   \eqref{tm}. This 
 reflects the fact that the spectrum of $M_L$ is fully embedded
 in that of  $M_{L+1}$.

\begin{thm} \label{thm:tm}
There exists a transfer matrix for the model  which can be expressed
by a recursion of order one. 
If one writes the transfer matrix from size $L-1$ to size $L$ by a block
decomposition of matrices of size $2^{L-1} \times 2^{L-1}$ as
\be \label{tmdecomp}
T_{L-1,L} = \left( \begin{array}{c}
T_1^{(L-1)} \\
T_2^{(L-1)}
\end{array} \right), 
\ee
then the matrix $T_{L,L+1}$ can be written as
\be \label{conjtm1}
T_{L,L+1} =   \left( \begin{array}{c}
T_1^{(L)} \\
T_2^{(L)}
\end{array} \right) 
\ee with 
\be \label{conjtm}
 T_1^{(L)} =   \left( \begin{array}{cc}
\ds T_{1}^{(L-1)} + \frac{1}{\alpha} T_2^{(L-1)} \,\,  & \,\, 
  \ds 2 T_2^{(L-1)} +  \frac{1}{\alpha}  T_2^{(L-1)}
(\sigma\otimes \id{L-2}) \\
\\
\ds T_2^{(L-1)}\,\,    & \,\,   \ds  \frac{1}{\alpha}  T_2^{(L-1)}
\end{array} \right)  \,\, \hbox{ and  } \,\,
T_2^{(L)} =  \left( \begin{array}{cc}
\ds 2T_2^{(L-1)}\,\,   & \,\,    \ds T_2^{(L-1)}  (\sigma\otimes \id{L-2}) \\ \\
\ds 0 \,\,     &\,\,   T_2^{(L-1)}
\end{array} \right).
\ee
This, along with the initial condition
\be \label{tmic}
T_{1,2} = \left( \begin{array}{c c}
1 + \beta + \alpha \beta & \alpha + \beta + \alpha \beta \\
\alpha & 1 \\
\alpha + \alpha \beta & \alpha \beta \\
0 & \alpha
\end{array} \right),
\ee
determines recursively  a family of transfer matrices.
\end{thm}

The proof  is  carried out  by  induction. We
first suppose that 
equation \eqref{tm} is satisfied for the transfer $L-1 \to L$.
 We then prove  \eqref{tm} for the transfer $L \to L+1$ using the 
conjectured formula \eqref{conjtm}  for the transfer matrices. For
convenience, we omit the subscripts denoting the dimension of the
identity matrices. Using the
decomposition \eqref{tmdecomp} and \eqref{mmdecomp} in the equation
$M_L T_{L-1,L} = T_{L-1,L} M_{L-1}$ yields  the following two identities:
\be \label{id1}
M_{L-1} T_1 - \alpha (\sigma \otimes \iden) T_1 + \alpha T_2 + \sigid
T_2 = T_1 M_{L-1},
\ee
and
\be \label{id2}
\alpha T_1 + M_{L-1} T_2 - T_2 - \alpha \sigid T_2 = T_2 M_{L-1}.
\ee
(Here, to simplify the notations we have written  $T_1$ instead of  $T_1^{(L-1)}$, 
$T_2$ instead of  $T_2^{(L-1)}$  and $\iden$  instead of $\id{L-2}$).
  We shall need  two other 
identities satisfied   by  the transfer matrices defined by the recursion
 \eqref{conjtm}. These identities  will be proved by 
induction.

 The third identity is the following:
\be \label{id3}
\alpha T_2 \sigid = \alpha T_1 - T_2.
\ee
 Assume that  this is true for  the blocks
$T_1,T_2$    of   the transfer
 matrix $T_{L-1,L}$  \eqref{tmdecomp}. Now, using \eqref{conjtm}, we have  for 
the corresponding matrices in   $T_{L,L+1}$
\be
\begin{split}
  \alpha T_1^{(L)} - T_2^{(L)} =   & \alpha
\begin{pmatrix}
\ds T_{1} + T_2/\alpha & \ds 2 T_2 +  T_2 (\sigma\otimes \id{L-2})/\alpha \\
\ds T_2 & \ds T_2/\alpha 
\end{pmatrix} - 
\begin{pmatrix}
\ds 2T_2 & \ds T_2  (\sigma\otimes \id{L-2}) \\
\ds 0 & T_2
\end{pmatrix}, \\
&=
\alpha
\begin{pmatrix}
\ds T_2   (\sigma\otimes \id{L-2}) & \ds 2 T_2 \\
\ds T_2 & 0
\end{pmatrix}
 = \alpha T_2^{(L)}  (\sigma\otimes \id{L-1}) \, ,
\end{split}
\ee
 which  proves  \eqref{id3}.

 The fourth   identity is
\be \label{id4}
T_2 [M_{L-1} ,\sigid] = [T_2,\sigid] \, . 
\ee
This equation is proved  by noting from  \eqref{mmdecomp}   that
\be
[M_{L-1} ,\sigid] = \begin{pmatrix}
  \sigid & \iden \\
  -\iden & -\sigid
\end{pmatrix} \, .
\ee
Then, using the induction hypothesis, we find
 that the left hand side of \eqref{id4} is given by 
\be
\begin{pmatrix}
  \ds 2T_2 & \ds T_2 \sigid \\
  \ds 0 & T_2
\end{pmatrix} 
\begin{pmatrix}
  \sigma \otimes \iden & \iden \\
  -\iden & -\sigma \otimes \iden
\end{pmatrix} =
\begin{pmatrix}
  T_2 \sigid & T_2 \\
  -T_2 & -T_2 \sigid
\end{pmatrix},
\ee
which is easily verified to be equal to 
 the right hand side of \eqref{id4}
  using again   the induction hypothesis.

 Finally,  using \eqref{conjtm} and \eqref{mmdecomp}, 
  we calculate  explicitly 
  $T_{L,L+1} M_L$ in square blocks of size $2^{L-1}$   as
\be \label{tmmm}
 \left( \begin{array}{c|c}
T_1 M_{L-1} + T_2 M_{L-1}/\alpha &
\alpha T_1 +2T_2 M_{L-1} - 2 T_2 + T_1 \sigid \\
- \alpha T_1 \sigid + 2 \alpha T_2 &
-2 \alpha T_2 \sigid + T_2 \sigid M_{L-1}/\alpha \\
\hline
T_2 M_{L-1} + T_2 &
\alpha T_2 - T_2/\alpha \\
 - \alpha T_2 \sigid & + T_2 M_{L-1}/\alpha \\
\hline
2T_2 M_{L-1} - \alpha T_2 \sigid & \alpha T_2 + T_2 \sigid + T_2
\sigid M_{L-1} \\
\hline
\alpha T_2 & T_2 M_{L-1} - T_2 - \alpha T_2 \sigid
\end{array} \right),
\ee
and similarly we write $M_{L+1} T_{L,L+1}$  as
\be \label{mmtm}
 \left( \begin{array}{c|c}
M_{L-1} T_1 + M_{L-1} T_2/\alpha &
T_2 +2 M_{L-1} T_2 - \alpha \sigid T_2 \\
- \alpha \sigid T_1 + 2 \alpha T_2 &
+ \sigid T_2/\alpha + M_{L-1} T_2 \sigid /\alpha \\
 & -\sigid T_2 \sigid \\
\hline
 M_{L-1} T_2 + T_2  & 
\alpha T_2 +  M_{L-1} T_2/\alpha \\
- \alpha \sigid T_2 & - T_2/\alpha -\sigid T_2 + T_2 \sigid\\
\hline
\alpha T_1 + 2 M_{L-1} T_2 -T_2 &
 2 \alpha T_2+ M_{L-1} T_2 \sigid\\
- 2 \alpha  \sigid T_2 & 
 + \sigid T_2  - \alpha \sigid T_2 \sigid\\
\hline
\alpha T_2 & M_{L-1} T_2 - T_2 - \alpha  \sigid T_2
\end{array} \right).
\ee
(Horizontal and vertical lines
  were inserted to  separate  
 the different blocks that compose  the matrices).

The fact that  the matrices \eqref{tmmm} and
\eqref{mmtm} are equal is a direct  consequence of 
 the four identities  \eqref{id1}, \eqref{id2},
 \eqref{id3} and \eqref{id4}. This
  completes  the proof of Theorem~\ref{thm:tm}.

There are many  properties of the matrix $T_{L,L+1}$ which can be
used to prove statements  about the steady state distribution of the
system.  
In particular, we  can use \eqref{nontriv} and the recursion \eqref{conjtm}
 to compute the kernel $|v_L \rangle$. The entries in $v_L$ are
necessarily polynomials if we start with the vector $|v_1 \rangle =
\begin{pmatrix} \alpha+\beta \\ \alpha \end{pmatrix}$. Because the
submatrix $T_2$ of $T_L$ is always proportional to $\alpha$, it is
easy to check that the value of $v_L$ for a configuration with $k$ 1's
is always proportional to $\alpha^k$.
In particular, the  last entry of  $v_L$  which corresponds to the configuration
with all sites occupied  is equal to   $\alpha^L$.

 We shall call the sum $Z_L$  of the 
  entries in  $v_L$, the  
{\em partition function}. This allows us to
 define the steady state probabilities as the vector
$|p_L \rangle = \frac 1{Z_L} |v_L \rangle$.
 Therefore  the probability of the configuration with all
sites occupied in a system of size $L$ is given by  $\alpha^L/Z_L$.
 This corresponds to the smallest probability. 
 Note that the polynomial  $Z_L=Z_L(\alpha,\beta)$  is the least
common multiple of the denominators of the entries of the kernel $p_L$
of $M_L$ provided  that the greatest common divisor  of the numerators of
the entries is one.
For example, the system of size one has
\be
|v_1 \rangle = \begin{pmatrix}
\alpha + \beta\\
\alpha 
\end{pmatrix},
\ee
whence $Z_1 = \alpha + 2\beta$. We find a remarkable property of the
partition function of the system, namely its super-extensive growth with 
the size of the system.

\begin{cor} \label{cor:den}
The partition function of the system of size $L$ is
given by
\be  \label{zeq}
Z_L = 2^{\binom{L-1}{2}} (1+2 \alpha)^{L-1} (1+\beta)^{L-1} (2 \alpha
+ \beta).
\ee
\end{cor}
This formula will be proved by induction. By definition of the 
 partition function we have 
\be
Z_{L+1} =  \langle 1_{L+1}|v_{L+1} \rangle 
  =  \langle 1_{L+1}| T_{L,L+1} | v_{L} \rangle,
\ee
where we have defined  $\langle 1_{K}|= \langle 1,\dots,1|$ to be the
line-vector of length  $2^K$ with all  entries  equal to 1, $K$
 being an arbitrary integer.
We now prove  by induction that $\langle 1_K|$ is a left-eigenvector 
 of $T_1^{(K)}$ and $T_2^{(K)}$.  More precisely, we show that
\be \label{t1t2}
\begin{split}
\langle 1_K| T_1^{(K)} &= 2^{K-1}(1+\alpha)(1+\beta) \langle 1_K|, \\
\langle 1_K| T_2^{(K)} &= 2^{K-1}\alpha(1+\beta) \langle 1_K|.
\end{split}
\ee
For the initial condition in \eqref{tmic}, we have  $\langle 1,1|
T_1^{(1)} = (1+\alpha)(1+\beta) \langle 1,1|$ and $\langle 1,1|
T_2^{(1)} = \alpha(1+\beta) \langle 1,1|$. 
Then,  using \eqref{conjtm} we express  $T_2^{(K)}$  in terms
of  $T_2^{(K-1)}$.  This leads to
\be
\langle 1_K|T_2^{(K)} = 2 \langle 1_{K-1}|T_2^{(K-1)}.
\ee
(Here, we have used the
 fact that  $\langle 1_{K-1}|$ is an eigenvector of 
 $T_2^{(K-1)}$ which
 in turn implies that    $\langle 1_{K-1}|T_2^{(K-1)} = 
 \langle 1_{K-1}|T_2^{(K-1)}  (\sigma\otimes \id{K-2})$).
 Next, we use 
\eqref{id3} to obtain $\langle 1_K|T_1^{(K)} = (1+1/\alpha) \langle
1_K|T_2^{(K)}$, thus proving the recurrence \eqref{t1t2}. 
Finally, we have 
\be
\begin{split}
Z_{L+1} &=  \Big( 2^{L-1}(1+\alpha)(1+\beta) \langle 1_L| +
  2^{L-1}\alpha(1+\beta) \langle 1_L| \Big) v_{L} \rangle, \\
&=2^{L-1}(1+2\alpha)(1+\beta) \langle 1_L|v_{L} \rangle \\
&=2^{L-1}(1+2\alpha)(1+\beta) Z_{L},
\end{split}
\ee
which, knowing $Z_1$, proves the desired formula \eqref{zeq}.

We now proceed by  calculating some  correlation functions in  the model:
\begin{cor} \label{cor:block}
For a system of size greater than $k+1$,
the probability of the first $k$ sites being occupied is
\be
 \label{eq:corrblock}
\langle \eta_1 \dots \eta_k \rangle =
\frac{\alpha^k}{2^{\binom{k}{2}} (1+2 \alpha)^k}.
\ee
\end{cor}
To prove this relation, we only have to consider   a system
of size $k+1$ as follows from
  Remark~\ref{rem:corrfn}. In this case, only two configurations contribute to
this expectation value --- either the last (i.e.the  $k+1$th) site  is
occupied or it is not. In the basis in which we have written our
Markov matrix, this corresponds to the last two entries of $|v_{k+1} \rangle$.

To obtain the sum of these two entries, we have to multiply the vector
$|v_{k+1} \rangle$ on the left by the vector $w_{k+1} = \langle
0,\dots,0,1,1|$ with 
$2^{k+1}-2$ zeros. Using \eqref{nontriv}, we have to calculate the
action of $\langle w_{k+1}|$ on the transfer matrix $T_{L-1,L}$. Note that
$w$  couples  only   with  the last two rows of the transfer matrix. From the
recursion \eqref{conjtm} and the initial condition \eqref{tmic}, we
see that the only nonzero entries in the last two rows are from the $2
\times 2$ block at the bottom right. Notice also that the column-wise
sum of these two blocks  is the same and is equal to $\alpha(1+\beta)$.
Therefore, we obtain
\be
\begin{split}
\langle \eta_1 \dots \eta_k \rangle &= \frac{\langle w_{k+1}|T_{k+1,k}|
  v_k \rangle}{Z_{k+1}} 
 = \frac{\alpha(1+\beta) \langle 0,\dots,0,1,1|  v_k \rangle}{Z_{k+1}} 
 = \frac{\alpha(1+\beta) Z_k \langle \eta_1 \dots \eta_{k-1} \rangle
  }{Z_{k+1}}  \\
&= \frac{\alpha}{2^{k-1} (1+2 \alpha)} \langle \eta_1 \dots \eta_{k-1} \rangle;
\end{split}
\ee
this recursion, along with the initial condition $\langle \eta_1
\rangle = \frac \alpha{1+2\alpha}$, proves \eqref{eq:corrblock}.

\section{Calculation of one-point and two-point correlations}
\label{sec:jw}

\subsection{Disorder Variables} 
 \label{subsec:dv}
We now introduce a set of non-local  variables that
 will allow us to calculate exact properties of the model. These are
 similar to the so-called interparticle distribution functions in
 \cite{AvrahamDoering}. 
 Consider  the random variable 
\be 
\xi_i =  (-1)^{\# \{1 \leq j \leq i| \eta_j = 1\}}
\ee
which takes values $\pm 1$ depending on whether the number of occupied
sites between  the first site  and the $i$th site
   is even or odd respectively. 
We  call $\xi_i$ the {\em disorder variable} at site $i$.
This definition is inspired from the 
theory of spin chains, where the  Jordan-Wigner transformation allows us
to write  fermionic creation and annihilation operators from spin
operators.
More precisely, let $S^+_j,S^-_j,S^z_j$ represent operators
which raise, lower and measure the spin at site $j$ respectively. They
satisfy the relations $\{S^+_j,S^-_j\}=\iden$ and
$[S^+_j,S^-_k]=0$. From these  operators one 
can construct fermionic operators $f_j,f^\dag_j$ which satisfy the
standard anticommutation relations 
\be
\begin{split}
f^\dag_j &= S^+_j (-1)^{\phi(j)},\\
f_j &= (-1)^{\phi(j)} S^-_j,
\end{split}
\ee
where $\phi(j) = \sum_{k=1}^{j-1}(1/2+S^z_k)$ measures the number of 
up-spins to the left of the site $j$. 
Our variable $\xi_i$ corresponds to  the variable $\phi(j)$ in the theory
of spin chains.

For an exclusion process like the one we consider here, $\xi_i$ has a more
convenient representation,
\be \label{jw}
\xi_i = \prod_{j=1}^i (1-2\eta_j).
\ee
 Note that these random variables satisfy $\xi_i^2=1$
since the occupation variables satisfy $\eta_i^2=\eta_i$.
The steady state expectation values of the $\xi_i$ variables satisfy
some remarkable properties, which we will state in this section. These
expectation values will be extremely useful in proving results for
the density and the evaporation rate.

\begin{thm} \label{thm:jwav}
For a system with size $L$ and $i<L$,
\be \label{jwav}
\langle \xi_i \rangle = \frac 1{(1+2\alpha)^i} \, ,
\ee
and
\be \label{jwbav}
\langle \xi_L \rangle = \frac \beta{(\beta+2\alpha) (1+2\alpha)^{L-1}} \, .
\ee
\end{thm}
 From \eqref{jw}, we have for all $i,1 \leq i \leq L$,
\be \label{jwrecur}
\xi_i = (1-2\eta_i) \xi_{i-1},
\ee
from which we also have
\be
\eta_i \xi_i = -\eta_i \xi_{i-1},
\ee
which substituting back in \eqref{jwrecur} gives
\be \label{jwrecur2}
\xi_i = \xi_{i-1} +2 \eta_i \xi_i.
\ee
We now write down  the evolution  equation for $\xi_i$ in the
bulk.  The main idea, and one of the reasons this variable is
 useful,  is that a transition  taking place strictly
between 1 and $i$ does not   affect the value of $\exi{i}$. The first bulk
transition 
in \eqref{bulk} does not change the number of particles, and the
second reduces it by two, and hence does not change the parity. We
will first write down the equation and then explain each term.
\be \label{jwss}
\begin{split}
0 = \frac d{dt} \exi{i} &= +\alpha \aver{-\xi_i} + \aver{\eta_i
  (-\xi_i)}-\alpha \aver{\xi_i} + \aver{\eta_i \xi_i}, \\
&= -2\alpha \aver{\xi_i} -2 \aver{\eta_i \xi_i}.
\end{split}
\ee
The first two terms describe ways of entering the configuration
contributing to $\exi{i}$. If one starts with  a configuration
contributing to 
$-\exi{i}$ and a particle enters or leaves  site 1 the parity will change
because of \eqref{lbound}. Similarly if one starts from configuration
contributing to $-\exi{i}$ and if the $i$th site is occupied
and that particle leaves that site, then the
parity changes. The last two terms 
describe ways of exiting the configuration contributing to
$\exi{i}$:  when a particle  enters from the left reservoir  or when a particle
 occupying site $i$ leaves it, we end up with a configuration
contributing to $-\exi{i}$. Using \eqref{jwss} in \eqref{jwrecur2},
one ends up with
\be
\exi{i} = \frac{\exi{i-1}}{1 + 2 \alpha},
\ee
which along with the initial condition, $\exi{0}=1$  leads us
to \eqref{jwav}. For the last site, the  balance equation is similar,
\be
0= \frac d{dt}\exi{L} = -2\alpha \exi{L} -2\beta\aver{\eta_L \xi_L}
\ee
because the last site exits with rate $\beta$ instead of rate one. And
using \eqref{jwrecur2}, one obtains
\be
\exi{L} = \frac{\beta \exi{L-1}}{\beta + 2 \alpha},
\ee
which leads to \eqref{jwbav}.

We now calculate  correlations among the disorder
variables. These are directly related to the quantities of interest in
the model. For example,
\be
\label{eq:disdens}
\aver{\xi_i \xi_{i-1}} = \aver{1-2\eta_i},
\ee
will  give  us the density.  We will also see the relation between  this
  problem and  a random walk  in two dimensions.
 For convenience, we introduce   a new variable
\be \label{ximn}
\xi_{m,n} = \frac{1-\xi_m \xi_n}2,
\ee
and define the expression $n_+$ to be the positive part of $n$. Namely
$n_+ = n$ if $n \geq 0$ and 0 otherwise.

\begin{thm} \label{thm:jwcorr}
For a system with size $L$ and $0 \leq n \leq m <L$,
\be \label{jwcorr}
\begin{split}
\aver{\xi_{m,n}} &= \frac \alpha{2^{(m+n-2)_+}(1+2\alpha)^m} \sum_{j=0}^{m-1}
  (1+2\alpha)^j 2^{(m-j-2)_+} 
 \sum_{k=1}^{m-n}
  \binom{(m+n-2)_+-(m-j-2)_+}{m-k},
\end{split}
\ee
and if $L>n>m$, then $\xi_{m,n} = \xi_{n,m}$.
\end{thm}

We first write down the  balance equation for the correlation
function for $m,n \geq 1$,
\be \label{jwcorrss}
\begin{split}
0=\frac d{dt}\aver{\xi_m \xi_n} &= -2 \aver{\eta_m \xi_m \xi_n} -2\aver{\eta_n \xi_m
  \xi_n}, \\
&= \aver{\xi_{m-1} \xi_n} + \aver{\xi_m \xi_{n-1}} -2 \aver{\xi_m \xi_n},
\end{split}
\ee
where we have used ideas very similar to \eqref{jwss} in the first
line and \eqref{jwrecur2} in the second line. In this  expression,
$\xi_0$ is set identically to 1.
The simplicity of the expression is due to the fact that the product
$\xi_m \xi_n$ is affected only by transitions taking place between the
$m$th and the $n$th sites. Any other transition either changes the
signs of both $\xi_m$ and $\xi_n$ (e.g, first site) or does not change the
sign of either of them (e.g, last site).
 From   equation \eqref{jwcorrss}, we deduce  the
recursion relation 
\be
\aver{\xi_{m,n}} = \frac{\aver{\xi_{m-1,n}}+\aver{\xi_{m,n-1}}}2.
\ee
  We now prove \eqref{jwcorr} by showing that it verifies
 this recursion and the boundary 
conditions. Since the formula \eqref{jwcorr} is valid in the
triangular region $0 \leq n \leq m$, we have to check that it is valid
for $m=n$ and for $n=0$. The case  $m=n$  is easily done: substituting $n=m$
in \eqref{jwcorr} gives    zero (because the sum is empty)
 and from \eqref{ximn},
we also obtain  $\xi_{m,m}=0$. For the case  $n=0$, we have 
\be
\begin{split}
\aver{\xi_{m,0}} =& \frac \alpha{2^{(m-2)_+}(1+2\alpha)^m} \sum_{j=0}^{m-1}
  (1+2\alpha)^j 2^{(m-j-2)_+} 
\sum_{k=1}^{m}
  \binom{(m-2)_+-(m-j-2)_+}{m-k},\\
=& \frac \alpha{2^{m-2}(1+2\alpha)^m} \left( (1+2\alpha)^{m-1}
  \sum_{k=2}^m \binom{m-2}{m-k} 
  +\sum_{j=0}^{m-2}
  (1+2\alpha)^j 2^{m-j-2} \sum_{k=m-j}^{m}
  \binom{j}{m-k} \right),
\end{split}
\ee
where we have split the sum according to whether $j=m-1$ or not, and
changed the limits of $k$ to count only the nonzero summands. The
summation on $k$ is easily done in both terms to give
\be
\aver{\xi_{m,0}} = \alpha \left( \frac 1{1+2 \alpha} +
\sum_{j=0}^{m-2} (1+2 \alpha)^{j-m} \right) = 
  \frac{(1+2\alpha)^m-1}{2(1+2\alpha)^m},
\ee
which is the correct answer knowing $\xi_{m,0} = (1-\xi_m)/2$ and the
expectation value $\exi{m}$ from Theorem~\ref{thm:jwav}.
Thus we have verified the boundary  conditions. We  now  need to verify the
recursion \eqref{jwcorrss}. We can take $m>n \geq 1$ which implies $m+n-3
\geq 0$. Then,
\be 
\begin{split}
\frac{\aver{\xi_{m,n-1}}+\aver{\xi_{m-1,n}}}2 &= \frac \alpha{2 \cdot 2^{m+n-3}} \sum_{j=0}^{m-1}
  (1+2\alpha)^{j-m} 2^{(m-j-2)_+}  \sum_{k=1}^{m-n+1}
  \binom{m+n-3-(m-j-2)_+}{m-k}\\
&+\frac \alpha{2 \cdot 2^{m+n-3}} \sum_{j=0}^{m-2}
  (1+2\alpha)^{j-m+1} 2^{(m-j-3)_+} 
 \sum_{k=1}^{m-n-1}
  \binom{m+n-3-(m-j-3)_+}{m-k-1}.
\end{split}
\ee
We  replace $j \to j+1$ in the second sum  to get the summands in
both these terms to be exactly the same except for the binomial
coefficients, in which the lower index differs by one. One can thus
combine both these sums using the usual addition formula for binomial
coefficients in the common range to get 
\be 
\begin{split}
& \frac{\aver{\xi_{m-1,n}}+\aver{\xi_{m,n-1}}}2 = \\   &\frac
\alpha{2^{m+n-2}} \sum_{j=1}^{m-1} (1+2\alpha)^{j-m} 2^{(m-j-2)_+} 
\left( \sum_{k=1}^{m-n-1}
  \binom{m+n-2-(m-j-2)_+}{m-k}   
  +\sum_{k=m-n}^{m-n+1}\binom{m+n-3-(m-j-2)_+}{m-k} \right)\\
&+\frac \alpha{2^{m+n-2}} (1+2\alpha)^{-m} 2^{m-2} \sum_{k=1}^{m-n+1}
\binom{n-1}{m-k}.  
\end{split}
\ee
 The second sum over $k$,  which   comes  from the contribution of  
  $\aver{\xi_{m,n-1}}$,  
involves  only two terms which can  again be summed using the addition
formula for binomial coefficients and the result can be included in the first
sum as the $k=m-n$ term.  The third  sum  over $k$ (which is
   also an  untouched term  from 
  $\aver{\xi_{m,n-1}}$  corresponding  to $j=0$)   is equal to 
  one because the binomial coefficient is nonzero
if $n-1 \geq m-k$, which is satisfied only for the upper limit,
$k=m-n+1$, where the value is one. Finally,   
 this  third sum over $k$ adds  to the
first sum as the $j=0$ term, giving the right hand side of
\eqref{jwcorr} and  confirming the recurrence relation.

\subsection{Density}
 We know
 from  Remark~\ref{rem:corrfn}  that the density at site
$k$ is a fixed quantity for sites $L>k$ and depends only on
$\alpha$. From \eqref{eq:disdens}, it is clear that $\eta_m =
\xi_{m,m-1}$. Using Theorem~\ref{thm:jwcorr} we arrive at an
explicit expression for the expectation value of the density,
\be
\begin{split}
\aver{\eta_m} &= \frac \alpha{2^{(2m-3)_+}(1+2\alpha)^m} \sum_{j=0}^{m-1}
  (1+2\alpha)^j 2^{(m-j-2)_+}  \binom{(2m-3)_+-(m-j-2)_+}{m-1}.
\end{split}
\ee
When $m=1$, we only have the $j=0$ term and the answer matches that
given by Corollary~\ref{cor:block} with $k=1$. For $m \geq 2$, we
split the sum according to  whether $j=m-1$ or not,
\be
\begin{split}
\aver{\eta_m} &= \frac \alpha{2^{2m-3}(1+2\alpha)^m} \left(
(1+2\alpha)^{m-1} \binom{2m-3}{m-1} + 
 \sum_{j=0}^{m-2}  (1+2\alpha)^j 2^{m-j-2} \binom{m-1+j}{m-1} \right).
\end{split}
\ee
We now multiply and divide this formula by two: the denominator in the
prefactor becomes $2^{2m-2}$ and the two terms inside the brackets are
multiplied by two. For the first term, we get
\be
2 \binom{2m-3}{m-1} =  \binom{2m-3}{m-2}+ \binom{2m-3}{m-1} =
\binom{2m-2}{m-1},
\ee
and for the second term, the power of two inside the summation becomes
$2^{m-j-1}$. We can then combine both these terms and write the
density explicitly as
\be
\dens{m} = \frac \alpha{2^{2(m-1)} (1+2\alpha)^m}\sum_{j=0}^{m-1}
\binom{m-1+j}{j} (1+2\alpha)^j 2^{m-1-j}. 
\ee
One can analyze the asymptotics of the density using Stirling's
formula and noting that the largest summand is the one where
$j=m-1$. One finds that
\be \label{densasymp}
\dens{m} \sim \frac 1{2 \sqrt{\pi m}},
\ee
{\em
  independent} of $\alpha$. This asymptotic result was obtained
 by  Lebowitz, Neuhauser and Ravishankar using stochastic coupling
 methods (see  Theorem 2 of \cite{JNR}). The  problem studied by these
 authors corresponds to the case  $\alpha=1$
 and  was  inspired by earlier studies of the 
  Toom model \cite{DLSS} on the semi-infinite lattice.

\subsection{Evaporation Rate}
Since, unlike the TASEP,  particles are {\it not} conserved in the bulk,
 a quantity like the current that is constant
 across all bonds does not exist in the present model.
 But we can define an evaporation rate at the 
bond $(m-1,m)$, as the correlation   $\langle \eta_{m-1} \eta_{m} \rangle$:
 this quantity does not depend on the size $L$ of the system 
 as long as $L>m$ because of Remark~\ref{rem:corrfn}.
 We  can calculate  the evaporation rate using the expansion
\be
\aver{\xi_{m,m-2}} = \aver{1-2\eta_m-2\eta_{m-1}+4 \eta_{m-1}\eta_m},
\ee
from which we have
\be
\aver{\eta_{m-1}\eta_m}=\frac{\aver{\xi_{m,m-2}}-1+2\aver{\eta_m}+
  2\aver{\eta_{m-1}}}4. 
\ee
An alternative  method is to use the balance  equation for the density
\be \label{densdb}
\frac d{dt} \dens{m} = \dens{m-1}-\dens{m} - 2 \aver{\eta_{m-1}\eta_m } =0,
\ee
to obtain
\be
\aver{ \eta_{m-1}\eta_m} =  \frac {\alpha^2}{2^{2m-3} (1+2\alpha)^{m}}
\sum_{i=0}^{m-2} \frac{m-1-i}{m-1} \binom{m-2+i}{i} \gamma^i 2^{m-2-i}.
\ee
 The asymptotics for the evaporation rate are given by 
\be \label{evapasymp}
\aver{\eta_{m-1}\eta_m} \sim \frac 1{8 \sqrt{\pi m^3}}.
\ee
One can similarly compute higher order correlations of a bunch of
consecutive sites $\aver{\eta_i \dots \eta_{i+k}}$ using
Theorem~\ref{thm:jwcorr}. The fact that the asymptotic evaporation
rate in \eqref{evapasymp} is given (upto a factor of 2) by the
derivative of the asymptotic density \eqref{densasymp} can probably be
explained by looking at the hydrodynamic equation for this model.

\subsection{Structure of  higher correlation functions}
 \label{sec:corrfn}

We outline some general properties of correlation functions in the
model. The reason one can make  strong statements about
correlations in this model is explained in 
Remark~\ref{rem:corrfn}. We will comment
on correlations for both order and disorder variables.

\subsubsection{Density correlations}
We earlier used the balance equation  for the density
\eqref{densdb} in computing the evaporation rate. We also found simple
closed form expressions for the correlations of sites $1,\dots,k$ being
occupied in Corollary~\ref{cor:block} and for correlations of the form
$\aver{\prod_{j=1}^k (1-2\eta_j)}$ in Theorem~\ref{thm:jwav}. In
Theorem~\ref{thm:jwcorr} we found a more complicated expression for
correlations of the form $\aver{\prod_{j=l}^k (1-2\eta_j)}$, where $l
\neq 1$ and $l<k$.
Here, we will say more about the  balance equations obeyed by more
general order variables. 

Consider a general correlation function of order $n$, written as  $\aver{\eta_{p_1}
  \cdots \eta_{p_n}}$. In general we know no closed form expression
for such a correlation function, but we can write down the
balance  equation satisfied by this  object. We group the positions
$i_1,\cdots i_n$ according to blocks of consecutive sites.
 The total  number of blocks is given by $j$ and  the
length of the $l$th block is  denoted by  $k_l+1$. The blocks are thus
labelled   as $(\eta_{i_1},\cdots,\eta_{i_1+k_1})$ up to
$(\eta_{i_j},\cdots,\eta_{i_j+k_j})$. Note that  neighbouring blocks are
separated by at least one site and that the size of a  block could be one.

\begin{thm}
The steady state   equation satisfied by the correlation function of
sites $i_1,\cdots, i_1+k_1,\cdots,i_j,\cdots,i_j+k_j$ where $i_1>1$ is given by
\be \label{dcorrs}
\begin{split}
n  \, \aver{\eta_{p_1} \dots \eta_{p_n}} = &\sum_{l=1}^j \Bigg\langle 
\left( \prod_{m=1}^{l-1} \eta_{i_m} \dots \eta_{i_m+k_m} \right) \\
&\eta_{i_l-1} (1-2\eta_{i_l}) \eta_{i_l+1} \dots \eta_{i_l+k_l}
\left( \prod_{m=l+1}^{j} \eta_{i_m} \dots \eta_{i_m+k_m} \right)
\Bigg\rangle.
\end{split}
\ee
\end{thm}

The main idea is to consider what happens for a single block
 made  of consecutive  $n$ sites $i,\dots,i+n-1$. The
 balance condition for such a  block   can be written easily,
\be \label{oneblock}
\begin{split}
0=\frac d{dt} \aver{\eta_i \dots \eta_{i+n-1}} =&
\aver{\eta_{i-1}(1-\eta_i)\eta_{i+1} \dots \eta_{i+n-1}} \\
&-\aver{\eta_{i-1}\eta_i\eta_{i+1} \dots \eta_{i+n-1}}
-n\aver{\eta_i \dots \eta_{i+n-1}}, \\
=& \aver{\eta_{i-1}(1-2\eta_i)\eta_{i+1} \dots \eta_{i+n-1}} \\
&-n\aver{\eta_i \dots \eta_{i+n-1}},
\end{split}
\ee
because the configuration with  all the sites $i,\dots,i+n-1$ occupied can
only be reached if the $(i-1)$th site is occupied, the $i$th site is
empty and the remainder are occupied; we can exit the configuration if
the $(i-1)$th site is occupied along with all the others from $i$ to
$i+n-1$ with rate one, and in $n$ different ways (each site between
$i$ and $i+n-1$ could jump) otherwise. The expression \eqref{oneblock}
is precisely 
\eqref{dcorrs} for $j=1$. Two blocks are separated by at
least one site and this argument applies independently to each
block. From each block $l$  we get a contribution of $k_l+1$ times
$\aver{\eta_{p_1} \dots \eta_{p_n}}$ all of which add to give the
factor $n$ on the left hand side of \eqref{dcorrs} whereas
 on  the right hand side  the first factor $\eta_{i_l}$
 of  each block $l$  is successively
 replaced by a factor $\eta_{i_l-1} (1-2\eta_{i_l})$
 whereas  the other blocks are left unchanged.

\subsubsection{Disorder variables correlations}
A nontrivial observation about one-point and two-point correlation
functions of the disorder variables $\xi_i$ is that their evolution 
equations \eqref{jwss} and \eqref{jwcorrss}, supplemented by
the recursion \eqref{jwrecur2}  are {\em closed} in the
sense that they involve only other one-point and two-point disorder
correlations respectively. In general, one would expect a hierarchy where the
 equations for lower correlations would necessarily
involve higher correlations making the problem extremely difficult to
solve for generic $n$-point correlations. As we will show below, this
property of decoupling, which is very  specific
to this model, holds for all $n$.

\begin{thm}
In the steady state, the correlation function of $n$ disorder
variables $\aver{\xi_{p_1} \dots \xi_{p_n}}$ satisfies the equation
\be
n \, \aver{\xi_{p_1} \dots \xi_{p_n}} = \sum_{k=1}^n \aver{\xi_{p_1}
  \dots \xi_{p_{k-1}} \xi_{p_k-1} \xi_{p_{k+1}} \dots \xi_{p_n}},
\ee
if $n$ is even, and 
\be
(n+2\alpha) \aver{\xi_{p_1} \dots \xi_{p_n}} = \sum_{k=1}^n \aver{\xi_{p_1}
  \dots \xi_{p_{k-1}} \xi_{p_k-1} \xi_{p_{k+1}} \dots \xi_{p_n}},
\ee
if $n$ is odd.
\end{thm}

 To prove these identities, one simply writes
  down the evolution  equation, which for even $n$  is given by 
\be
 \frac d{dt}  \aver{\xi_{p_1} \dots \xi_{p_n}} = -2\sum_{k=1}^n
\aver{\xi_{p_1} 
  \dots \xi_{p_{k-1}} \xi_{p_k} \eta_{p_k} \xi_{p_{k+1}} \dots
  \xi_{p_n}},
\ee
 and for odd $n$  is given by 
\be
\frac d{dt}  \aver{\xi_{p_1} \dots \xi_{p_n}} = -2\alpha
\aver{\xi_{p_1} \dots \xi_{p_n}} -2\sum_{k=1}^n \aver{\xi_{p_1} 
  \dots \xi_{p_{k-1}} \xi_{p_k} \eta_{p_k} \xi_{p_{k+1}} \dots
  \xi_{p_n}} \, . 
\ee
 One has to treat separately  the even and  the odd cases  because a
particle entering from 
the left reservoir  makes a difference to the product $\xi_{p_1}
\dots \xi_{p_n}$  only  if $n$ is odd. The terms inside the sum itself are
easily explained analogous to \eqref{jwss} and \eqref{jwcorrss}; each
factor $\eta_{p_k}$ occurs because the jump of the particle at site
$p_k$ changes only the sign of $\xi_{p_k}$ and keeps all the others
intact. Then  we use the  recursion \eqref{jwrecur2} for each term in the
sum to prove the result.
 We emphasize  that no problem arises  if any of the $p_k$'s are
consecutive. If $p_{k-1} = p_k-1$ for a particular value of $k$, 
a factor  $\xi_{p_{k-1}}^2 =1 $  appears in the expectation
value and drops out from  the product. As an example, the correlation
function for three consecutive sites  satisfies:
$ (3+2\alpha) \aver{\xi_i \xi_{i+1} \xi_{i+2}} = \aver{\xi_{i-1}
  \xi_{i+1} \xi_{i+2}} + \aver{\xi_i} + \aver{\xi_{i+2}}.$

\section{Spectrum of the Markov matrices}
 \label{sec:mm}

 In this section,  we mention some spectral
properties of the  Markov matrix  of the asymmetric annihilation
 model.   We observed that its 
  characteristic polynomials  factorize into
  linear factors and can be written explicitly. We first
  define the polynomials $A_L(x)$ and $B_L(x)$ as
\be
\begin{split}
A_L(x) &= \prod_{k=0}^{\lceil L/2 \rceil} (x+2k)^{\binom{L-1}{2k}}, \\
B_L(x) &= \prod_{k=0}^{\lfloor L/2 \rfloor}
(x+2k+1)^{\binom{L-1}{2k+1}}.
\end{split}
\ee

\begin{conj}
The characteristic polynomial $P_L(x)$ of $M_L$ is given by 
\be \label{charpoly}
P_L(x) = A_L(x) A_L(x+2\alpha+\beta) B_L(x+\beta) B_L(x+2\alpha),
\ee
and successive ratios of characteristic polynomials are given by
\be
\frac{P_{L+1}(x)}{P_L(x)} = B_L(x+1) B_L(x+2\alpha+\beta+1) A_L(x+\beta+1)
A_L(x+2\alpha+1). 
\ee
\end{conj}

If this is true, the Markov matrix has only $2L$ distinct eigenvalues.
 In particular, the 
  negative of all the factors in the denominator $Z_L$ are   roots of
 the characteristic  polynomial;  for example, the denominator $Z_3$
according to   Corollary~\ref{cor:den} is $2 (1+2\alpha)^2
  (1+\beta)^2(2\alpha+\beta)$ whereas $P_3(x)$ is given by 
\be
x \left( x+2 \right)  \left( \beta+x+1 \right) ^{2} \left( 2\,\alpha+x+1
\right) ^{2} \left( 2\,\alpha+2+x+\beta \right)  \left(
  2\,\alpha+x+\beta \right). 
\ee

We also conjecture that the Markov matrices are {\em maximally
  undiagonalizable} in the sense that each eigenvalue seems to have
exactly one eigenvector independent of the number of times the
eigenvalue appears
 (this implies  that the Jordan blocks are of the maximum possible size).
 In other words,  the degeneracy seems to be  so strong that the
Markov matrix $M_L$ which is of size $2^L$ has only $2L$ eigenvectors. 
  We can also write $P_L(x)$ in a different manner. We   define $n_1(m)$,
for any integer $m$, as $(-1)$ raised to the number of ones in the
binary expansion of $m$. For example, $n_1(1) = n_1(2)=-1, n_1(3)=1$.
 
\begin{conj}
Let us  index the rows and columns from 0 to $2^L-1$. The
characteristic polynomial of $M_L$ is then given by 
\be \label{conj2}
P_L(x) = \prod_{i=0}^{2^L-1} (x-(M_L)_{(i,i)}- \alpha n_1(i)).
\ee
\end{conj}
 Equation~\eqref{charpoly} has been checked for systems of length $\le 7$.
 The proof of this conjecture seems to be a nontrivial problem
 in determinant evaluation \cite{tewodros}.

We have been able to prove a much weaker result. Consider the matrix
$\widetilde M_L = M_L- \alpha \sigid$ written in the block diagonal
decomposition as 
\be
\left( \begin{array}{c|c}
M_{L-1} - \alpha (\sigma \otimes \id{L-2}) &  (\sigma
\otimes \id{L-2}) \\
0 & M_{L-1} - \id{L-1} - \alpha (\sigma \otimes \id{L-2} )
\end{array} \right),
\ee
using \eqref{mmdecomp}. Notice that the two diagonal blocks are
$\widetilde M_{L-1}$ and $\widetilde M_{L-1}- \id{L-1}$. Using
\eqref{mmic}, one can see that $\widetilde M_1$ is upper
triangular. Therefore $\widetilde M_L$ is also upper triangular for
all $L$. Therefore the eigenvalues of $\widetilde M_L$ are simply the
elements on the diagonal. Using the special nature of the diagonal
blocks, one can easily prove by induction that 
the characteristic polynomial of $\widetilde M_L$ is given by
\be
\widetilde P_L(x) = \prod_{k=0}^{L-1} (x+\alpha+k)^{\binom{L-1}k}
(x+\alpha+\beta+k)^{\binom{L-1}k}.
\ee
Note  the close similarity with the characteristic polynomial of
$M_L$ in \eqref{charpoly}.

\section{Discussion and Conclusion} \label{sec:con}

   In this work we have studied a 
 nonequilibrium  system on a finite size lattice with
 open  boundaries  in which particles diffuse and  interact
 through hard-core exclusion and pairwise annihilation. 
 The breaking of detailed balance in the bulk of the
 system  is ensured by the asymmetric
 hopping rules and by the absence of pair-creation process. Besides,
 the  difference in the chemical potentials  of the left and
 the right reservoirs is also a source  of nonequilibrium behaviour.
  The bulk dynamics is characterized by a single dimensionless 
  parameter $\lambda$
  which represents the ratio  between   evaporation
 and   hopping rates \eqref{bulk}. For  $\lambda=0$, the model
 is identical to the totally asymmetric exclusion process
 with open boundaries \cite{DEHP}. In the
 present work we have derived exact results for the case  $\lambda=1$.
 The fact that  $\lambda=1$ is a special point  can be
 understood if one writes the Markov matrix $M_L$ of this stochastic
 process as a non-Hermitian spin chain operator  using the Pauli matrices
 \cite{Stinchcombe1,Rittenb}. For general values of $\lambda$ we obtain
\be
 M_L =  \sum_{i=1}^{L-1} M_{i,i+1} + R + L \,,
\ee
where
\begin{eqnarray}
 \label{operatorspinchain}
  M_{i,i+1} &=& S^+_iS^-_{i+1} + \lambda  S^+_iS^+_{i+1}
 + \frac{ 1 -\lambda}{4} S^z_iS^z_{i+1} + \frac{ 1 +\lambda}{4} S^z_i
-  \frac{ 1 -\lambda}{4} S^z_{i+1} - \frac{ 1 +\lambda}{4} \, , \nonumber \\
     L  &=&    \alpha \left( S^-_1 + \lambda  S^+_1
 - \frac{1-\lambda}{2} S^z_1 - \frac{1+\lambda}{2} \right) \, , \nonumber \\
  R &=& \beta \left(  S^+_L -\frac{1 - S^z_L}{2} \right) \, .
\end{eqnarray}
We recall that the spin operators are given by:
$ S^+ = \begin{pmatrix}
  0 & 1 \\
  0 & 0
  \end{pmatrix} ,  \,\,\,
S^- = \begin{pmatrix}
  0 & 0 \\
  1 & 0
  \end{pmatrix}   $ and
$ S^z = \begin{pmatrix}
  1 & 0 \\
  0 & -1
  \end{pmatrix} .$ 
  For $\lambda=1$, we observe that all the interaction terms of the
  type $S^z_iS^z_{i+1}$ disappear from the spin chain operator: this
  corresponds to the {\it free fermion point.} This absence of
  interaction gives an alternative explanation for the simplicity of
  the model at the special value $\lambda=1$. However, we remark that
  at this special point,  the  boundary terms give a nonlocal
  contribution  because of lack of
  periodicity and therefore the usual  strategies for diagonalizing the
  quadratic Hamiltonian on a periodic ring (i.e. Fourier Transform
  or  Bogoliubov transformation) do not seem to apply here. The transfer
 matrix technique allows us to by-pass these  difficulties.
 For $\lambda=0$ the spin chain operator
  \eqref{operatorspinchain} represents the totally asymmetric
  exclusion process which is also exactly solvable. It would therefore
  be of interest to explore the integrability properties of this
  system for general values of $\lambda$ taking into account the
  boundary conditions.  Another possible extension is to allow
  backward hopping of the particles (which corresponds to the
  partially asymmetric case) and to formulate the boundary conditions
  so that the model remains solvable.
 
 For this asymmetric annihilation process, we have been able to derive
 exact combinatorial expressions for correlation functions such as
 local densities and evaporation rates.  The use of disorder variables
 has been helpful.  However, the main tool that we have introduced is
 a recursion relation between systems of two consecutive sizes. This
 recursion is encoded in a semi-similarity operator (the transfer
 matrix Ansatz) that  conjugates 
 the  Markov matrices for  systems of sizes $L$ and $L+1$. 
 The model studied here admits a transfer matrix Ansatz whereas the
 TASEP does not (as we explicitly checked on small systems).
 Conversely, the TASEP can be solved using a quadratic matrix product
 representation (which allows the calculation of physical observables such
 as the density and two-point correlation functions) whereas the
 steady state weights of the asymmetric annihilation process cannot
 be written easily as a simple matrix product (see Appendix~\ref{sec:rec}).
 Indeed, the fact that the normalization $Z_L$ grows super-extensively
 as $2^{L(L-1)/2}$ (see \eqref{zeq}) implies that the matrices for the
 stationary weights should involve $L(L-1)/2$ tensor products as in
 the case of multi-species exclusion processes \cite{Pablo,Sylv1} and
 the calculation of physical observables would be a true challenge.

  To summarize, the method for solving this model differs considerably
  from that used for the exclusion process.  The transfer matrix
  Ansatz encodes in a particularly efficient way recurrence relations
  for the steady state probabilities and it allows one to deduce
  information about the steady state in a rather elementary manner.
  We believe that the existence of a transfer matrix Ansatz is rather
  general: it applies to the multi-species exclusion process on a ring
  \cite{Aritaetal} and perhaps in a non-obvious manner to the ASEP
  with open boundaries.
 
  Of a more fundamental interest is the following question.  The
  system is out of equilibrium and therefore its steady state violates
  detailed balance: this implies the existence of elementary currents
  between microscopic configurations.  However, due to the evaporation
  of particles, there is no obvious way of defining a conserved
  current in the system. It would be interesting to find an observable
  that demonstrates at the macroscopic level the breaking of detailed
  balance.
 
 Finally, we remarked above that the normalization $Z_L$ grows
 super-exponentially with the size of the system for $\lambda=1$
 whereas for $\lambda=0$, $\log Z_L$ is extensive in $L$. Although
 $Z_L$ has no direct physical interpretation, it grows with $L$
  roughly as the ratio of the most probable configuration to the
 least probable one.  It would be of interest to see when this
 transition from exponential to super-exponential growth occurs as
 $\lambda$ varies and to interpret it physically.  More generally, a
 challenging problem is to calculate for arbitrary values of $\lambda$
 the scaling function $f(\lambda)$ defined by $Z_L \sim
 \exp(f(\lambda) L^2)$.  Related questions were addressed in rice-pile
 models by P. Pradhan and D. Dhar \cite{Dhar} and in the so-called
 Raise and Peel model by de Gier and collaborators \cite{degieretal}.

\section*{Acknowledgements}
 We thank   C.~Godr\`eche and J.~M.~Luck for
 discussions at the beginning of the project and we are grateful
 to  O.~Golinelli,  P. Krapivsky, D. Dhar and R. Rajesh for useful comments.
  We  also thank S. Prolhac for his help in  the initial stages of this work
 and S. Mallick for a careful reading of the manuscript.
  The first author (A.A.) would  also like to
 thank D.~Zeilberger, J.~L.~Lebowitz and E.~R.~Speer for discussions
 and for hospitality 
 at the Mathematics Department of Rutgers University. 
\appendix

\section{Recursions for the
 Steady State Probabilities}
 \label{sec:rec}
 
 In this appendix, we give an algorithm that allows to calculate
  recursively the steady state weights of configurations in a system
  of size $L$ knowing the ones for size $L-1$.  These recursions are
  based on numerical observations and are conjectural. They are stated
  in this appendix in order to be compared with the recurrences for
  the TASEP. They could also be used as a tool to build a matrix
  product representation.

 We define the {\em pushing operator} $P$ that acts on a block of 1's
followed by a block of 0's. $P(1^j 0^k)$ is given by a linear
combination of all possible $\binom{j+k}k$ binary words with $j$ 1's
and $k$ 0's.  The coefficient of a given word is a power of 2 as
follows: each move of the 1 which began at position $j$ (i.e. the
rightmost 1) contributes a factor of 2; each move of the 1 which began
at position $j-1$ contributes a factor of 1/2; each move of the 1
which began at position $j-2$ contributes a factor of 2 and so on.  We
give some examples,
 \be \label{pex}
\begin{split}
P(100) &=  (100)+2(010)+4(001), \\
P(1100) &= (1100)+2(1010)+4(1001) 
 +(0110)+2(0101)+(0011).
\end{split}
\ee
In the first example, there was only one 1 and we gained a factor of
two for every push of the 1. In the second example, we  gained a
factor of two for  each push of the rightmost  1 but lost a factor of
two for   each  push of the other  1.

  We  shall now construct  an operator $N$ which acts on words 
  and  generates  a linear combination of words  of  size one less.
   The operator  $N$ will be  defined in  two stages: 
  first we  specify  the action of $N$
   on  special  words that  we call {\em sub-configurations}
 (which are building blocks for general words),  then
 we explain how to extend this  action  on arbitrary words by concatenation. 

 A sub-configuration is a word that contains a single block of empty
sites. In a {\em bulk sub-configuration,} of the type $10^k1$ with
$k>0$, the block is surrounded by an occupied site on the left and on
the right. In a {\em right (left) sub-configuration,} of the type
$10^k$ ($0^k1$) with $k>0$, there is a single occupied site on the
left (right). The configuration with all empty sites will be treated
separately.  We shall distinguish three cases:

\begin{itemize}
\item{Bulk sub-configuration:} 
\be \label{bulksub}
N(10^k1) = \sum_{j=0}^{\lfloor k/2 \rfloor} 2^j P(1^{2j+1} 0^{k-2j}).
\ee

\item Left sub-configuration:
the   relation  involves  $\alpha$,
\be \label{lbsub}
\begin{split}
N(0^k 1) = \sum_{j=0 }^{\lfloor (k-1)/2 \rfloor} 2^{j}
P(1^{2j+1} 0^{k-1-2j})
 + \alpha\sum_{j=0}^{\lfloor k/2 \rfloor}  2^{j} P(1^{2j}
0^{k-2j}) .
\end{split}
\ee

\item Right sub-configuration:
the  relation  now  involves $\beta$,
\be \label{rbsub}
\begin{split}
N(1 0^k) =  \beta \sum_{j=0 }^{\lfloor (k-1)/2 \rfloor}  2^{j}
P(1^{2j+1} 0^{k-1-2j}) 
 +  (1+\beta)\sum_{j=0}^{\lfloor k/2 \rfloor} 
2^{k-1-j} P(1^{2j} 0^{k-2j}) .
\end{split}
\ee

\end{itemize}

We are now in a position to construct  the operator $N$  for any
configuration except the one with  all empty
sites. Any nonempty configuration can be written as $\tau = 0^{k_0}
1^{l_1} 0^{k_1} \dots  1^{l_m} 0^{k_{m}}$. In this notation, a bulk
sub-configuration is $10^{k_i}1$ for $i=1,\dots,m-1$, the left
sub-configuration is $0^{k_0}1$ and the right sub-configuration is
$10^{k_m}$ assuming $k_0$ and $k_m$ to be  nonzero. If either of them is
zero, the corresponding boundary sub-configuration does not exist. The
idea is to  use the algorithms for each of these
  sub-configurations as defined in
\eqref{bulksub}, \eqref{lbsub} and \eqref{rbsub}  and
concatenate.
  The actual formula depends on whether $k_0$ is positive or not:
\be \label{confrec}
N(\tau) = \begin{cases}
  (0^{k_0}1) \oplus    {\ds \bigoplus_{i=1}^{m-1}}\left( 1^{l_i-1} \oplus (10^{k_i}1)
  \right) \oplus 1^{l_m-1} \oplus (10^{k_m}), & \\
  & \hspace{-1cm} \text{if } k_0>0, \\
  \alpha {\ds \bigoplus_{i=1}^{m-1}} \left(1^{l_i-1} \oplus (10^{k_i}1)
  \right) \oplus 1^{l_m-1} \oplus (10^{k_m}), & \\
  & \hspace{-1cm} \text{if } k_0=0, 
  \end{cases}
\ee
where the symbol $\oplus$ denotes concatenation
and where we use the summation formulae for the sub-configurations defined
previously. We emphasize  that the size of the configuration is exactly
one less than $\tau$. Note also that the right sub-configuration term
$N(10^{k_m})$ 
is empty if $k_m=0$. For example,
\be
\begin{split} \label{ex:decomp}
N(010) &= ( (1)+\alpha  (0)) \oplus (\beta  (1) + (1+\beta)  (0)), \\
&= \beta  (11) + \alpha \beta  (01) + (1+\beta)  (10) + \alpha
(1+\beta)  (00).
\end{split}
\ee
We have thus defined the operator $N$ for all words except  those
 consisting only of 0's. The  operator $N$ encodes 
  recursions which are precisely those given by  the transfer matrix Ansatz
 as we checked on system of size $\le 7$.

 The configuration with all empty sites requires a slightly different
  algorithm: this is the only case where the recursion is different
  from the one provided by the transfer matrix in $\eqref{conjtm}$. In
  fact, $N(0^L)$ involves all $2^{L-1}$ configurations of size $L-1$
   and the coefficients depend on whether these smaller
  configurations end in a one or a zero.  We need here a related but
  different pushing operator, $P'$, which we define presently.
  $P'(1^j 0^k)$ is again a sum of all possible $\binom{j+k}{k}$
  configurations but the coefficient (which is again a power of two)
  is assigned differently. We first assign $2^k$ to the configuration
  $1^j 0^k$ and this time, we divide by   two for every push of
  the rightmost  1,  multiply  by   two for every push of the 
  second-from-right 
  1, divide by two  for the third-from-right 1 and so on. For example, \be \label{ppex}
\begin{split}
P'(110) &= 2(110) + (101) + 2(011) \\
P'(1100) &= 4(1100)+2(1010)+(1001) +4(0110)+2(0101)+4(0011).
\end{split}
\ee
(Note  the similarities  and differences  between \eqref{pex} and \eqref{ppex}.)
The recursion for the configuration with all zeros is given by
\be \label{probzero}
\begin{split}
N(0^L) &= \beta \left( \sum_{j=0 }^{\lfloor (L-3)/2 \rfloor} 2^{j}
P(1^{2j+1} 0^{L-3-2j}) \oplus 1 
 +  \alpha \sum_{j=0}^{\lfloor (L-2)/2 \rfloor}  2^{j} P(1^{2j}
0^{L-2-2j}) \oplus 1 \right) \\
&+ (1+\beta) \left( \alpha \sum_{j=0 }^{\lfloor (L-3)/2 \rfloor} 2^{j}
P'(1^{2j+1} 0^{L-3-2j}) \oplus 0 
 +   \sum_{j=1}^{\lfloor (L-2)/2 \rfloor}  2^{j-1} P'(1^{2j}
0^{L-2-2j}) \oplus 0 \right) \\
&+2^{L-2} (1+\alpha)(1+\beta) (0^{L-1}).
\end{split}
\ee 
We remark that  the powers of two are governed by the two
pushing operators $P$ and $P'$.
 Besides,  except for
the exceptional term with all zeros, there is a prefactor  $\beta$ if
the configuration ends in one and a prefactor  $1+\beta$ if it ends in
zero. 
 
 Equations \eqref{confrec} and \eqref{probzero} fully define the action
 of $N$ on an arbitrary  binary word. The operator $N$ acting on a word
 of size $L$ returns a linear combination of words of size $L-1$.
 If  ${\mathcal C}$ is  a configuration of the system of size $L$, then
 its unnormalized weight   $W({\mathcal C})$ is defined as
   \be
    W({\mathcal C}) = \langle {\mathcal C} |  v_{L} \rangle \,,
    \ee
    where  $v_{L}$  is the steady state vector constructed using
    the transfer matrix Ansatz \eqref{constructvL}. 
   We claim that  $W({\mathcal C})$ can be calculated knowing the weights
 of the  configurations of size $L-1$ from the following formula:
 \be 
 W({\mathcal C}) = W(N({\mathcal C})) \, .
 \ee
 Namely, the weight of  ${\mathcal C}$
 is the linear combination  of the weights of the 
  configurations of smaller size  generated by applying the operator
 $N$ to ${\mathcal C}$. For example, from \eqref{ex:decomp}, we deduce
\be
W(010) = \beta  W(11) + \alpha \beta  W(01) + (1+\beta)  W(10) + \alpha
(1+\beta)  W(00).
\ee
  
 A challenging problem would be to find a matrix  representation (or more generally
 an algebra) that embodies the reduction rules  \eqref{confrec} and \eqref{probzero}.
 We believe that  such an algebra does exist as  in the case  of ASEP.
  However, we emphasize that we have bypassed the  matrix product  representation
  altogether thanks to  the transfer matrix Ansatz defined in \eqref{tm}.

\end{document}